\documentstyle[12pt,aasms4]{article}

\slugcomment{Submitted to Astrophysical Journal (Letters).}

\lefthead{Scoville et al.}
\righthead{Imaging of Arp 220}

\newbox\grsign \setbox\grsign=\hbox{$>$} \newdimen\grdimen 
\grdimen=\ht\grsign
\newbox\laxbox \newbox\gaxbox
\setbox\gaxbox=\hbox{\raise.5ex\hbox{$>$}\llap
     {\lower.5ex\hbox{$\sim$}}}\ht1=\grdimen\dp1=0pt
\setbox\laxbox=\hbox{\raise.5ex\hbox{$<$}\llap
     {\lower.5ex\hbox{$\sim$}}}\ht2=\grdimen\dp2=0pt

\def\kms    {\ifmmode{{\rm ~km~s}^{-1}}\else{~km~s$^{-1}$}\fi}

\def\Lsun     {~$L_{\odot}$}
\def\Msun     {~$M_{\odot}$}
\def\deg      {{\ifmmode^\circ\else$^\circ$\fi} } 

\def\h2     {H$_2$}
\def\etal   {{\sl et~al.~}}
\newbox\grsign \setbox\grsign=\hbox{$>$} \newdimen\grdimen 
\grdimen=\ht\grsign
\newbox\simlessbox \newbox\simgreatbox
\setbox\simgreatbox=\hbox{\raise.5ex\hbox{$>$}\llap
     {\lower.5ex\hbox{$\sim$}}}\ht1=\grdimen\dp1=0pt
\setbox\simlessbox=\hbox{\raise.5ex\hbox{$<$}\llap
     {\lower.5ex\hbox{$\sim$}}}\ht2=\grdimen\dp2=0pt

\begin{document}

\title{NICMOS Imaging of the Nuclei of Arp 220}

\author{N. Z. Scoville,~~A. S. Evans}
\affil{California Institute of Technology, Pasadena, CA 91125}

\author{N. Dinshaw}
\affil{UCO/Lick Observatory, University of California, Santa Cruz, CA 95064}

\author{R. Thompson,~~M. Rieke,~~G. Schneider,~~F.J. Low,~~D. Hines,~~B. Stobie}
\affil{Steward Observatory, University of Arizona, Tucson, AZ 85721}

\author{E. Becklin}
\affil{University of California, Los Angeles, CA 90095}

\and 
\author{H. Epps}
\affil{UCO/Lick Observatory, University of California, Santa Cruz, CA 95064}



\begin{abstract}

We report high resolution imaging of the ultraluminous infrared galaxy Arp
220 at 1.1, 1.6, and 2.22 $\mu$m with NICMOS on the HST.  The
diffraction-limited images at 0.1--0.2\arcsec\ resolution clearly resolve
both nuclei of the merging galaxy system and reveal for the first time a
number of luminous star clusters in the circumnuclear envelope. The
morphologies of both nuclei are strongly affected by dust obscuration,
even at 2.2 $\mu$m : the primary nucleus (west) presents a crescent shape,
concave to the south and the secondary (eastern) nucleus is bifurcated by
a dust lane with the southern component being very reddened. In the
western nucleus, the morphology of the 2.2 $\mu$m emission is most likely
the result of obscuration by an opaque disk {\it{embedded}} within the
nuclear star cluster. The morphology of the central starburst-cluster in
the western nucleus is consistent with either a circumnuclear ring of star
formation or a spherical cluster with the bottom half obscured by the
embedded dust disk. Comparison of cm-wave radio continuum maps with the
near-infrared images suggests that the radio nuclei lie in the dust disk
on the west and near the highly reddened southern component of the eastern
complex.  The radio nuclei are separated by 0.98\arcsec\ (corresponding to
364 pc at 77 Mpc) and the half-widths of the infrared nuclei are
$\sim$0.2-0.5\arcsec\ . At least 8, unresolved infrared sources --
probably globular clusters -- are also seen in the circumnuclear envelope
at radii 2-7\arcsec\ . Their near-infrared colors do not significantly
constrain their ages.

\end{abstract}


\section{Introduction}

Arp 220 (IC 4553/4), with an infrared luminosity of 1.5 x 10$^{12}$
\Lsun~at $\lambda$=8-1000 $\mu$m, is one of the nearest ultraluminous
infrared galaxies  ({\cite{soi87}}). Visual wavelength images reveal two
faint tidal tails, indicating a recent tidal interaction (cf.
{\cite{joe85}}), and high resolution ground-based radio and near-infrared
imaging show a double nucleus ({\cite{ban87}}, {\cite{gra90}}). The radio
nuclei are separated by 0.98\arcsec\ at PA $\sim 90\deg$
({\cite{ban95}}).  The system is also extraordinarily rich in molecular
gas ($M_{H_2}\sim9\times10^9$\Msun, Scoville, Yun \& Bryant 1997) and
about 2/3 of this gas and dust is concentrated in a thin disk with radii
$\leq 250$ pc.  Ground-based observations at 10 to 30 $\mu$m suggest that
the far-infrared luminosity originates from a region of similar size
({\cite{wyn93}}). To power the energy output seen in the infrared by young
stars requires a star formation rate $\sim10^2$\Msun~yr$^{-1}$.
Alternatively, if the luminosity originates from an AGN, this source must
be sufficiently obscured by dust that even the mid-infrared emission lines
are highly extinguished since spectroscopy with ISO shows no evidence of
very high ionization lines at wavelengths out to 40$\mu$m
({\cite{stu96}}).

In this letter we report near infrared imaging at $\lambda$=1-2.2$\mu$m
with 0.1--0.2\arcsec\ resolution obtained using the NICMOS camera on the
HST. These high resolution data clearly reveal the complex nature of the
nuclei with the morphologies of both near-infrared nuclei strongly
affected by dust obscuration and enable an improved registration of the
infrared relative to the radio nuclei. Throughout this letter, we adopt a
distance of 77 Mpc (1\arcsec\ corresponds to 373 pc) for Arp 220 using
H$_0$=75 \kms Mpc$^{-1}$ ({\cite{soi87}}).

\section{Observations and Results}

Arp 220 was observed using the NICMOS camera 2 on the HST on 4 April 1997
as part of the Early Release Observations (ERO) program. Camera 2 has a
256$\times$256 HgCdTe array with 0.076\arcsec\ pixels, providing a
19.2\arcsec\ field of view. Images were obtained in the F110W, F160W and
F222M filters, providing resolutions of 0.11, 0.16 and 0.22\arcsec\,
respectively. For both the galaxy and an HST guide star, observed to
determine the point-spread-function (PSF), a four step dither pattern was
executed with offsets of 5.4 pixels (0.41\arcsec ). At each position in
the dither pattern on Arp 220, two integrations, each lasting 128 sec,
with multiple non-destructive reads (MULTIACCUM) were obtained. The data
were dark subtracted, flat fielded and corrected for cosmic rays using the
IRAF software NICMOS-pipeline routines. The dithered images were shifted
and coadded using the drizzle routine in IRAF.  The flux scale employed
calibration factors of 2.35$\times$, 2.77$\times$, and
6.11$\times10^{-6}$~Jy (ADU/sec)$^{-1}$ at 1.1, 1.6, and 2.22 $\mu$m
({\cite{rie97}}). The measured fluxes at H (1.6$\mu$m) and K (2.22$\mu$m)
for the nuclear sources and for a 5\arcsec\ aperture agree to better than
10\% with those obtained from ground-based imaging by Carico \etal (1990).
At 1.1$\mu$m a direct comparison is not possible since the standard J
filter is at 1.25$\mu$m. The rms noise levels in the final images are 12,
15, 35 $\mu$Jy (arcsec)$^{-2}$ at 1.1, 1.6, and 2.22 $\mu$m,
respectively.

For observations with NICMOS camera 2, the angular resolutions are a
factor of two lower at 2.22$\mu$m than at 1.1$\mu$m. This can seriously
hinder attempts to analyze color gradients where there are compact
emission sources unless the images at each wavelength are convolved or
deconvolved to a common resolution. We have adopted two approaches :
convolving the two short wavelength images to the resolution of the raw
image at 2.22$\mu$m and deconvolving the two long wavelength images to the
resolution at 1.1$\mu$m. The former is, of course, the safest approach but
provides more limited spatial resolution. For the latter approach,
deconvolution was done using the Richardson-Lucy algorithm without damping
({\cite{bus97}}) to achieve 0.14\arcsec\ resolution.

In Figure 1 (color plate) the images for the entire camera 2 field from
the 3 bands are combined in false color. The image clearly reveals the
structure around the twin nuclei separated by $\sim$ 1\arcsec\ E-W in
addition to extended emission from the system out to $\geq$8\arcsec. At
least eight unresolved ($\leq 30$pc) sources are also seen at radii
2--7\arcsec\ -- the brightest being approximately 5\arcsec\ southwest of
the nucleus.  All of them have bluer colors than the nuclei but their
magnitudes are sufficiently bright that they cannot be individual stars.
We hereafter refer to them as clusters. The brightest of these clusters
(1, 2, and 4) are seen in the visible band HST imaging of Shaya \etal
(1994) and they are also clearly seen in subsequent I-band HST imaging
(Borne and Lucas 1997).

The central region is shown in contour form in Figures 2-4 (Plates 2-4)
using : the original drizzled images at each wavelength; images convolved
to 0.22\arcsec$\,$ resolution; and deconvolved to 0.11\arcsec$\,$
resolution, respectively.  In each case, the lower right panel shows the
ratio of the 2.22$\mu$m/1.1$\mu$m data.  Coordinate offsets in Figures 2-4
are measured from the peak of the deconvolved 2.22$\mu$m emission in the
west nucleus. Flux measurements for the nuclear components and all
identified cluster sources are listed in Table 1 along with measured sizes
(FWHM) for resolved structures.

\section {Analysis and Discussion} 

Most of the near infrared emission from the nuclear region of Arp220 must
be stellar (rather than hot dust or shock-excited emission lines) since
the strength of the CO overtone absorption at 2.35$\mu$m requires that
$\geq$90\% of the observed near-infrared flux in a 5\arcsec\ aperture
arises from stars ({\cite{arm95}}, {\cite{shi96}}).  In previous
ground-based images (eg. {\cite{gra90}}), the core of Arp 220 appears as
two nuclei separated E-W by 0.95\arcsec\ . In the HST-NICMOS images, the
eastern nucleus breaks up into northern (NE) and southern (SE) components
separated by 0.4\arcsec\ -- the latter appears highly reddened and is
barely visible in the 1.1$\mu$m image. And, the bright nucleus on the west
(W), is seen as a {\it {crescent}} arc in the new images. Thus, for both
nuclei, their morphology, even at $\lambda$=2.2$\mu$m (where the
extinction  is 10 times less than A$_V$) appears strongly influenced by
dust extinction. In addition, a new component is seen $\sim
0.6$\arcsec~~south of the western nucleus (S) and another midway between
the two nuclei.

\subsection {Registration}

The accuracy of the guide star position for the Arp 220 observations was
insufficient to independently determine the absolute positions of the
near infrared images. We have therefore registered the infrared images by
comparison with cm-wave radio continuum maps for which the coordinates are
well-determined. Graham \etal (1990) matched the brightest radio sources
(A and B, {\cite{ban95}}) with the brightest portions of the eastern and
western nuclei (i.e. components NE and W).  However, this placement now
seems unlikely since the separation and position angles of the two radio
components and the two brightest infrared components are significantly
different. Instead, we suggest that both of the two brightest radio
components are located within the areas of very high extinction -- south
of the W component and between the NE and SE sources. This registration
places the weak radio source C precisely on the southern (S) infrared
source which is relatively blue and therefore unlikely to be strongly
extincted.  Using the 2.22$\mu$m/1.1$\mu$m ratio image (Figure 3) to
estimate the relative extinction across the region, we have crudely
corrected the 2.22$\mu$m image for extinction under the assumption that
the dust is in the foreground, i.e. not mixed with the stars. In Figure 5,
the radio source contours from Baan \& Haschick (1995) are shown
superposed on this `extinction-corrected' 2.22$\mu$m image using the
registration discussed above.  Radio source B lies within the dark area
below the crescent on the W source while A is at the north of the reddened
SE infrared source. Support for our registration of the near infrared
emission is provided by the strong morphological similarity of the
infrared structure shown here and the H$_2$CO emission ({\cite{ban95}}) --
the latter exhibits similar offsets relative to the radio continuum.

\subsection {Reddening}

The 2.22$\mu$m/1.1$\mu$m ratio images in Figures 2-4 may be used to probe
variations in the extinction assuming that the intrinsic color of the
underlying stars is not strongly varying. The false color image for the
deconvolved data (Plate 1) clearly shows an overall increase in the
reddening towards the central regions of Arp 220 -- in the annulus at
2-4\arcsec\ radii, m$_{1.1-2.2}$=2.30 mag, while inside 2\arcsec\  radius
, m$_{1.1-2.2}$=3.19 mag.  In the central 2\arcsec\ , Figure 3 reveals
even larger variations : high reddening is seen over the SE source where
the color (1.1-2.2$\mu$m) is 3.86 mag and in a nearly complete ring
surrounding the bright W source where the color is 3.31 mag.  For a
population of stars with ages $\leq 10^7$ yrs, typical colors are J-H =
-0.2, H-K $\sim$0.0 mag. For a standard extinction law ({\cite{rie85}}),
and the young population colors, the derived visual extinctions are A$_V$=
17 (W), 15(S), 24(SE), 18(NE), and 13(All) mag based on the observed H-K
magnitudes given in Table 1. For an old population (J-H $\sim$0.74, H-K
$\sim$0.24 mag), the visual extinctions are 13, 11, 20, 14, and 9 mag
respectively, similar to the 1\arcsec\ resolution estimates obtained by
Larkin \etal (1995) from Pa$\alpha$ and Br$\gamma$ line ratios.
Considerably higher estimates were obtained by Sturm \etal (1996) and it
is clear that there are undoubtedly areas in the nucleus from which
virtually no radiation escapes, even at 2.2$\mu$m.

\subsection {Nuclear Structure}

The complex near infrared structure of Arp 220 is due to there being
multiple centers of star formation activity (as in Arp 299, cf.
Wynn-Williams \etal 1991) and strongly varying dust obscuration within the
merger nuclei.  The relatively simple double-nuclei structure of the radio
continuum ({\cite{ban95}}) and the mm-wave dust continuum ({\cite{sco97}})
suggest that the major mass concentrations are associated with the W and
SE sources. The NE and S sources would then be relatively minor mass
concentrations, albeit with high present-epoch luminosity.

If the intrinsic, {\it{unobscured}} shapes of the expected stellar
emission are either ellipsoidal (in the case of the nuclear core
clusters), point-like (for unresolved stellar clusters) or disk-like (for
young stars formed in disk), the unobscured morphology should appear
spherical or elliptical depending on the viewing angle.  For both the E
and W nuclei, the morphology at 1.1-2.2$\mu$m is clearly not elliptical :
the western nucleus appears as a crescent and the eastern nucleus is split
into two components (NE and SE) with very different colors. Given our
suggested registration of the radio and infrared, it is evident that the
true centers of the nuclei are so obscured that neither is seen directly
at 2.2$\mu$m.

The crescent or partial-ring morphology of the western component is
clearest in the deconvolved 2.2$\mu$m image (Figure 4), but it is also
apparent in the original 2.2$\mu$m data (Figure 2). This morphology might
readily arise under two circumstances : if there is an obscuring disk of
dust (and gas) embedded within the {\it{spheroidal}} nuclear star cluster
or if a central starburst {\it{ring}} or disk of young stars is partially
obscured by its own dust.  Figure 6 models the surface brightness
distribution for a star cluster in which the midplane is cut by a totally
opaque dust disk; here, the observed crescent-shape arises due to an
increase in geometrical path length through the cluster periphery.  The
second hypothesis with a ring or torus of massive star formation is
similar to models proposed for Seyfert galaxy nuclei (eg. Maolino \& Rieke
1995, Genzel \etal 1995).

Two important characteristics of the observed emission should be
emphasized : the crescent shape and the extremely abrupt cutoff by dust to
the south of the W source.  The former strongly implies that the dust is
{\it{embedded}} in the core cluster/disk rather than being a foreground
dust lane; the latter implies that the dust is in a extremely thin, but
very opaque disk. This geometry agrees with that suggested by Scoville,
Yun \& Bryant (1997) based on arcsec resolution imaging of the mm-CO
line.  The low velocity dispersion of the molecular gas implied a very
thin disk ($\Delta$z=16 pc). A similar disk structure in the nuclei has
been suggested by Baan \& Haschick (1995) based on their H$_2$CO data.

The morphology of the eastern nucleus is not so distinctive, yet the
apparent bifurcation of the nucleus into lowly and highly reddened
components (NE and SE respectively) and the probable positioning of the
radio nucleus in an area of weak 2.2$\mu$m emission both suggest a nuclear
dust disk -- like that in the W nucleus. For both nuclei, the highest
extinctions occur to the south of the radio centers, suggesting that the
dust disks are tilted such that the south is the near side.

The measured infrared sizes (FWHM) of the W, SE and NE source components
are 0.2-0.5\arcsec\ , corresponding to 75-185 pc. The separation of the
radio nuclei (A and B) is 0.98\arcsec\ or 364 pc in projection. Their
actual separation is $\sim$600 pc if the geometry suggested by Baan \&
Haschick (1995) and Scoville, Yun \& Bryant (1997) is correct since the
nuclei would then be well off the major axis of their orbits.  Within each
of the nuclei, Scoville, Yun \& Bryant (1997) detected CO emission
profiles of full-width $\sim$250 \kms. If the rotation velocities of gas
within each nuclear cluster is $\sim$125 \kms ~at 75 pc radius, their
dynamical masses are $\sim3\times10^8$ \Msun .

\subsection {Unresolved Clusters}

 Magnitudes and positions of eight clusters for which reliable
measurements could be made are listed in Table 1 in order of decreasing
flux at 1.1 $\mu$m.  All of these sources are too bright to be individual
stars.  Given our resolution ($\sim$ 30 pc), the fact that they are
unresolved is entirely consistent with their being globular clusters.
Their measured offsets from the 2.2$\mu$m centroid imply projected
galactic radii of 0.9--2.7 kpc. The apparent clustering of these sources
around Arp 220, with none seen beyond 7.5\arcsec\ radius, suggests that
most of them must be associated with Arp 220 rather than in the
foreground. Given the fact that we have measured only two colors and must
use one color to constrain the reddening, the remaining color does not
strongly discriminate the ages of these clusters. The cluster sources
outside 2\arcsec\ radius have measured colors at 1.1, 1.6, and 2.2$\mu$m
which are consistent (within the uncertainties) with those of globular
clusters in NGC 5128 ({\cite{fro84}}), indicating that they could have age
$\geq 10^9$ yrs and essentially no near-infrared reddening (A$_V\leq$
5mag).  The absolute K magnitude of the brightest cluster (M$_K$= -13.5
mag) in Arp 220 is $\sim$1.5 magnitudes brighter than the brightest
cluster in NGC 5128 ({\cite{fro84}}) and is comparable in brightness to
the tip of the globular cluster luminosity function for ellipticals
({\cite{ajh94}}).  
very young 
much 
that massive, globular clusters 
event in the Arp 220 (cf. {\cite{fal85}}).

\acknowledgements

The NICMOS project has been supported by NASA grant NAG 5-3042. It is a
pleasure to acknowledge assistance and discussions of L. Armus, L.
Bergeron, C. Boone, J. Goldader, A. Fruchter, J. Jensen, T. Soifer, J.
Surace, and R. White. We are grateful to K. Borne and R. Lucas for sharing
their I-band image of Arp 220 with us.

\clearpage

%
%

\def\aas     {A\&AS}

\vfill\eject

\figcaption[fig1.cps]{False color image of the 1.1, 1.6, and 2.22$\mu$m
data for the entire field of view (19\arcsec\ ). The long wavelength
observations were deconvolved to 0.14\arcsec\ resolution.  \label{fig1}}

\figcaption[fig2.ps]{Contour and grey-scale images for the 1.1, 1.6, and
2.22$\mu$m data for the central 2.5\arcsec\ are shown together with the
ratio 2.22$\mu$m/1.1$\mu$m using the original-resolution drizzled images.
The coordinates are offsets in $\alpha$ and $\delta$ from the peak at
2.22$\mu$m ($\alpha_{1950}$= 15$^h$32$^m$46$s\atop .$90, $\delta_{1950}$=
+23$^\circ$40$^\prime$07$^{\prime\prime}_.$94). The contours are spaced
logarithmically by factors of 1.16, 1.17, and 1.16 down from the peak on
the W source at 5.1, 15, and 20 mJy (arcsec)$^{-2}$ at 1.1, 1.6 and
2.22$\mu$m. In the ratio map, the peak value is 15.8 and the contours are
spaced by factors of 1.54.  \label{fig2}}

\figcaption[fig3.ps]{Similar to Figure 2 except that the short wavelength
observations were convolved to the a resolution of 0.22\arcsec ( (i.e. the
original resolution at 2.22$\mu$m). The contours are spaced
logarithmically by factors of 1.13, 1.14, and 1.16 down from the peak on
the W source at 3.2, 11, and 19 mJy (arcsec)$^{-2}$ at 1.1, 1.6 and
2.22$\mu$m.  In the ratio map, the peak value is 15.2 and the contours are
spaced by factors of 1.34.  \label{fig3}}

\figcaption[fig4.ps]{Similar to Figure 2 except that the long wavelength
observations were deconvolved to 0.14\arcsec\ resolution using
Richardson-Lucy algorithm without damping. The contours are spaced
logarithmically by factors of 1.21, 1.22, and 1.26 down from the peak on
the W source. In the ratio map, the contours are spaced by factors of
1.65.  \label{fig4}}

\figcaption[fig5.ps]{Radio continuum sources at $\lambda$=6 cm are
superposed as contours on the `extinction-corrected' 2.22$\mu$m image (see
text). The radio source structure is derived from the deconvolved source
sizes and PA's given by Baan \& Haschick (1995). The registration of the
radio and infrared is discussed in the text. The weak radio component C
coincides with the infrared source S which is not apparent in the
corrected 2.22$\mu$m image due to its low extinction, ie. relatively blue
color. The coordinates are offsets from the peak at 2.22$\mu$m
($\alpha_{1950}$= 15$^h$32$^m$46$s\atop .$90, $\delta_{1950}$=
+23$^\circ$40$^\prime$07$^{\prime\prime}_.$94).  \label{fig5}}

\figcaption[fig6.ps]{A galactic-nucleus star cluster with an embedded,
opaque dust disk provides a possible explanation of the observed
morphology in the Arp 220 W source (and possibly the NE-SE complex). In
the model the stellar density is constant out to the cutoff radius and the
disk is inclined 20\deg to the line-of-sight.  \label{fig6}}

\vfill\eject

\begin{deluxetable}{lrrcccccccc}
\footnotesize
\tablewidth{0pc}
\tablecaption{Source Properties}
\tablehead{
\colhead{ Component } 		&
\colhead{ $\Delta \alpha^a$ } 	&
\colhead{ $\Delta \delta^a$ } 	&
\colhead{ m$_{1.1}^b$ }		&
\colhead{ m$_{1.25}^c$ }        &
\colhead{ m$_{1.6}^b$ }    	&
\colhead{ m$_{2.2}^b$ }		&
\colhead{ m$_{1.25-1.6}$  }	&
\colhead{ m$_{1.6-2.2}$  }	&
\colhead{ FWHM }  	        &
\colhead{ Aperture       }	}
\startdata
W           &   0.00 &   0.00  & 16.29 & 15.40 & 14.10 & 13.04 & 1.30 & 1.06 & $0\farcs49\times 0\farcs22$ & $0\farcs92\times 0\farcs58$ \nl
S           & E 0.21 & $-0.65$ & 18.00 & 17.28 & 16.09 & 15.16 & 1.19 & 0.93 & $0\farcs26\times 0\farcs19$ & $0\farcs40\times 0\farcs40$ \nl
SE          & E 1.00 & $-0.33$ & 18.70 & 17.68 & 16.31 & 14.83 & 1.37 & 1.48 & $0\farcs26\times 0\farcs23$ & $0\farcs40\times 0\farcs40$ \nl
NE          & E 1.13 &   0.07  & 17.47 & 16.57 & 15.27 & 14.14 & 1.30 & 1.13 & $0\farcs34\times 0\farcs26$ & $0\farcs55\times 0\farcs55$ \nl
All sources & E 0.57 & $-0.24$ & 13.72 & 13.08 & 11.95 & 11.13 & 1.13 & 0.82 &                             & 5\arcsec circular           \nl
Clusters:   &        &         &       &       &       &       &      &      &                             &                             \nl
~~~~~1	    & W 1.47 & $-4.80$ & 20.50 & 20.15 & 19.34 & 19.12 & 0.81 & 0.22 & 3\farcs7,4\farcs3,5\farcs6  & 4\farcs6,5\farcs4,7\farcs0  \nl
~~~~~2	    & E 4.76 & $-2.36$ & 21.35 & 21.03 & 20.25 & 20.21 & 0.78 & 0.04 &         \arcsec             &       \arcsec               \nl
~~~~~3	    & W 0.63 & $-2.35$ & 21.71 & 21.16 & 20.11 &  ---  & 1.05 & ---  &         \arcsec             &       \arcsec               \nl
~~~~~4	    & W 5.17 &   4.56  & 21.81 & 21.56 & 20.89 &  ---  & 0.66 & ---  &         \arcsec             &       \arcsec               \nl
~~~~~5	    & E 1.29 & $-4.46$ & 22.14 & 21.70 & 20.77 & 20.70 & 0.93 & 0.07 &         \arcsec             &       \arcsec               \nl
~~~~~6	    & E 3.83 & $-1.06$ & 22.48 & 22.19 & 21.47 &  ---  & 0.72 & ---  &         \arcsec             &       \arcsec               \nl
~~~~~7	    & W 0.58 &   4.74  & 22.87 & 22.52 & 21.71 &  ---  & 0.81 & ---  &         \arcsec             &       \arcsec               \nl
~~~~~8	    & E 7.11 &   1.83  & 22.89 & 22.32 & 21.25 &  ---  & 1.07 & ---  &         \arcsec             &       \arcsec               \nl
\enddata
\hskip -6mm
\tablenotetext{a}{Offsets in \arcsec\ relative to the peak at
2.2\micron, $\alpha_{1950} = 15^{\rm h}32^{\rm m}46\fs90$, $\delta_{1950} =
+23^\circ40^\prime07\farcs94$ based on the registration of the infrared
and radio as discussed in the text. \vskip 1mm}
\tablenotetext{b}{Magnitudes based on the flux calibration of
Rieke et al.\ (1997) and have typical uncertainty of $\pm0.1$~mag. m = 0
corresponds to 1922, 1075, and 667 Jy at 1.1, 1.6, and 2.22$\mu$m. \vskip 1mm }
\tablenotetext{c}{Magnitudes interpolated from m$_{1.1}$ and m$_{1.6}$ magnitudes.}
\end{deluxetable}

%
%
%
%
%

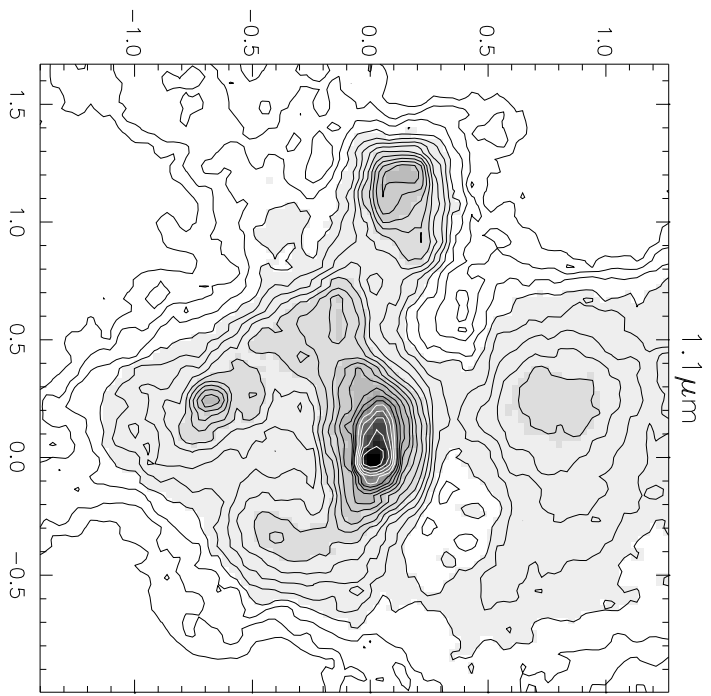
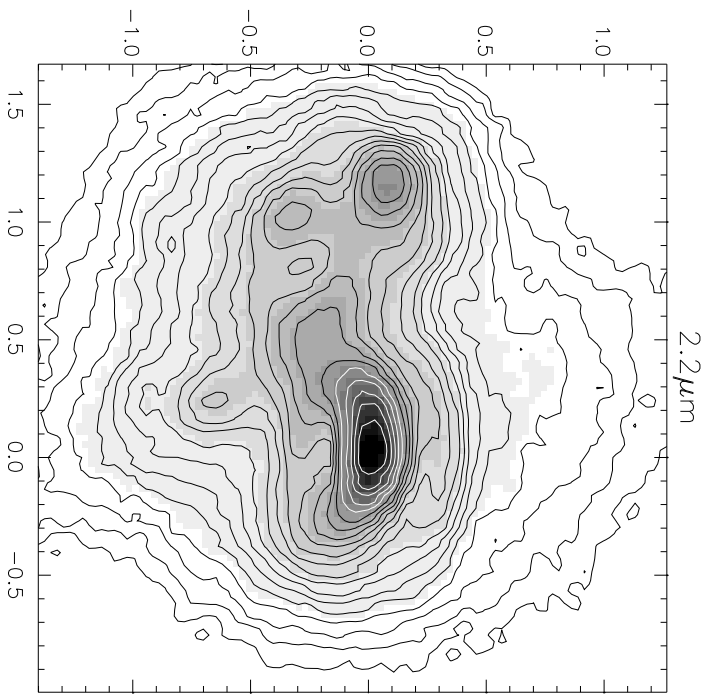
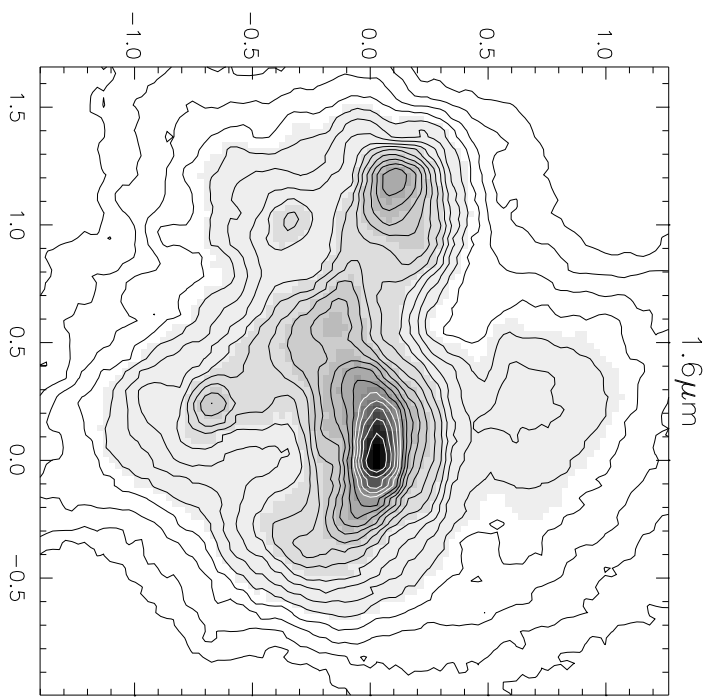
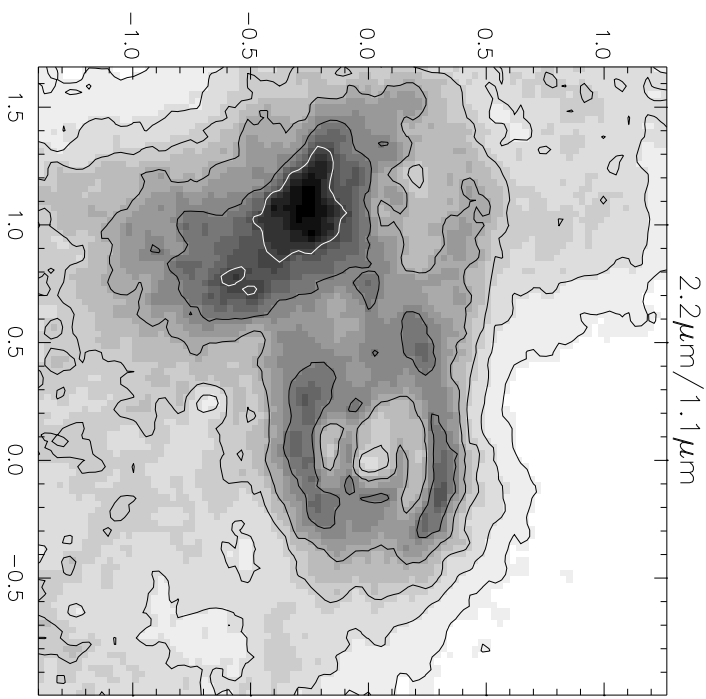
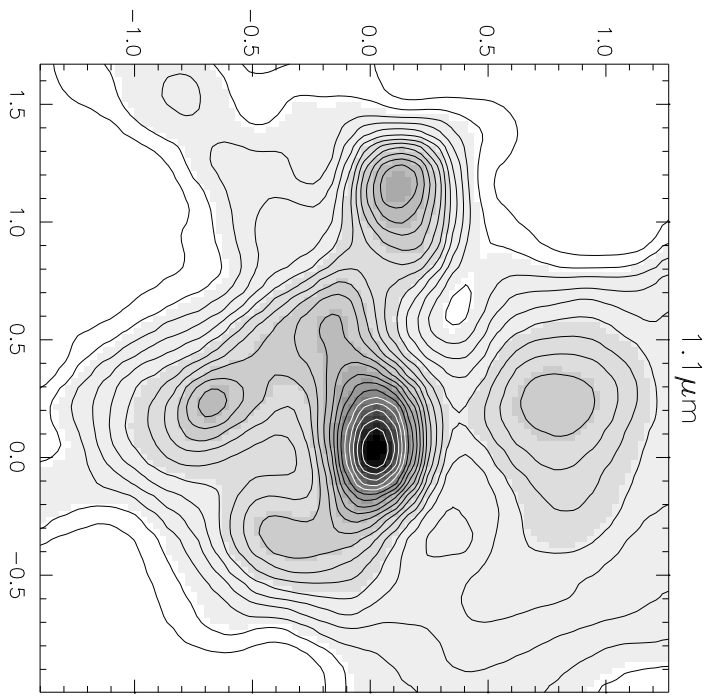
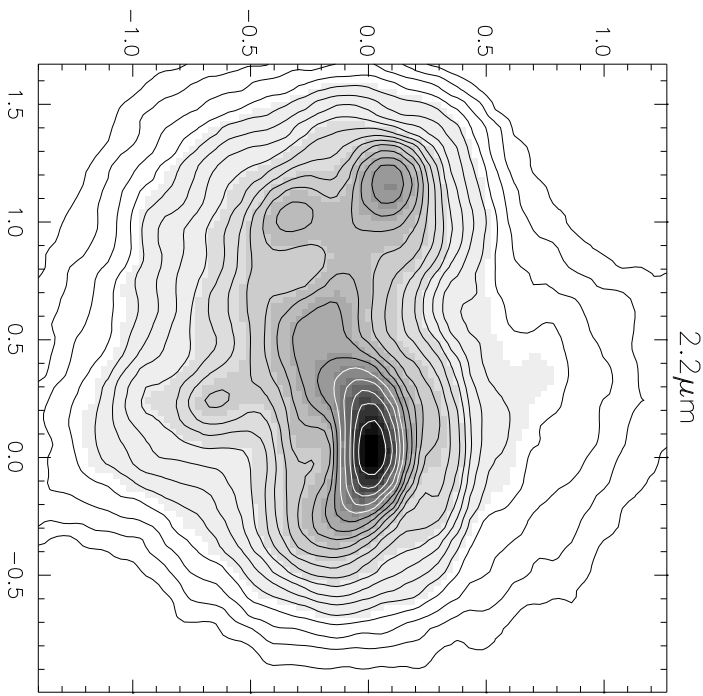
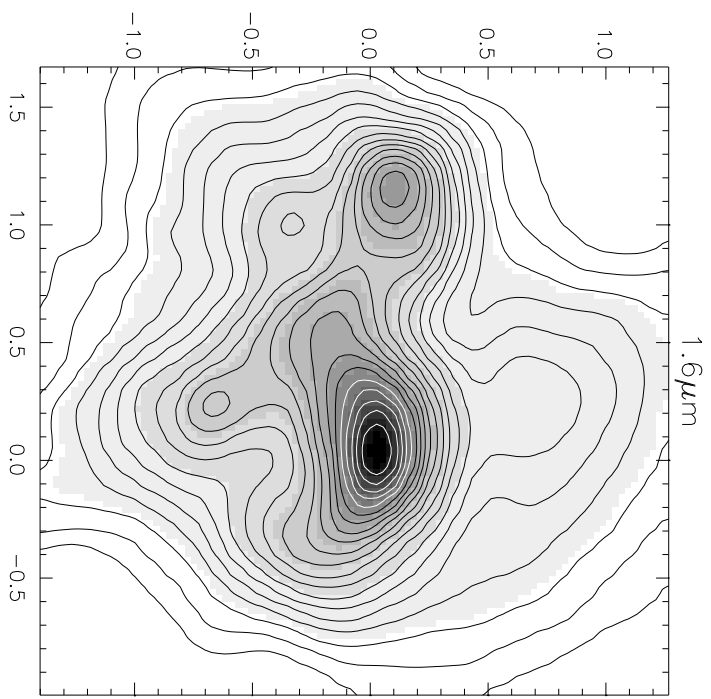
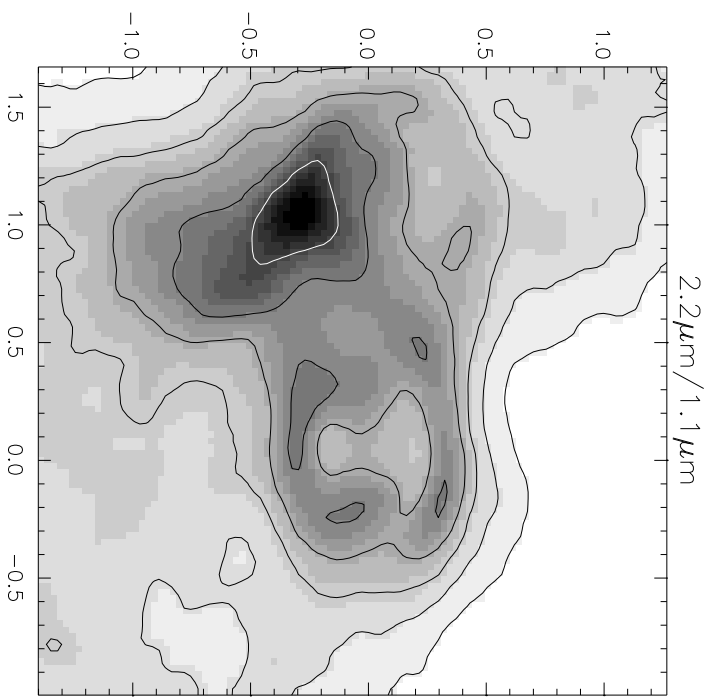
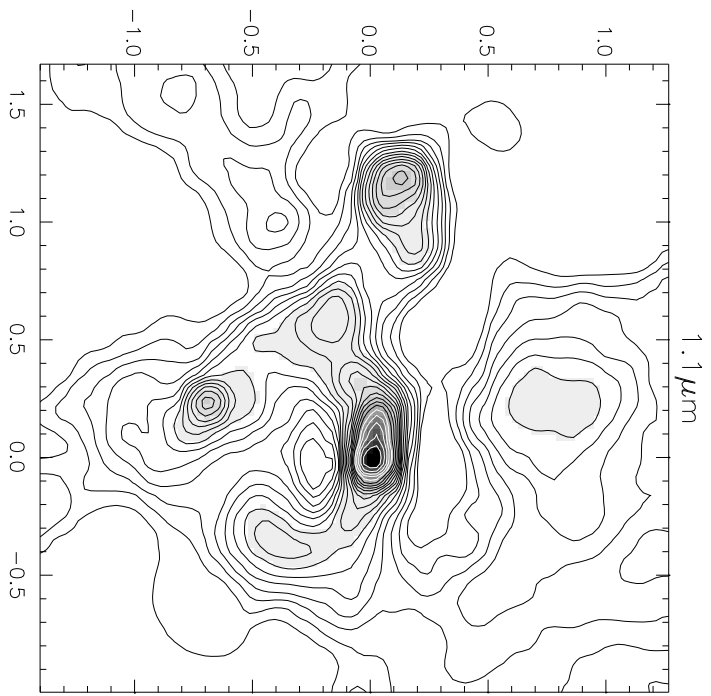
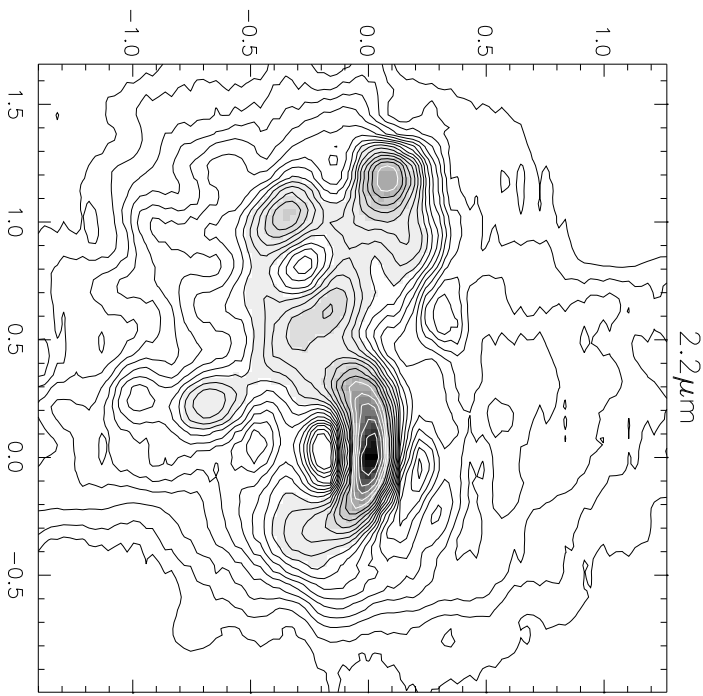
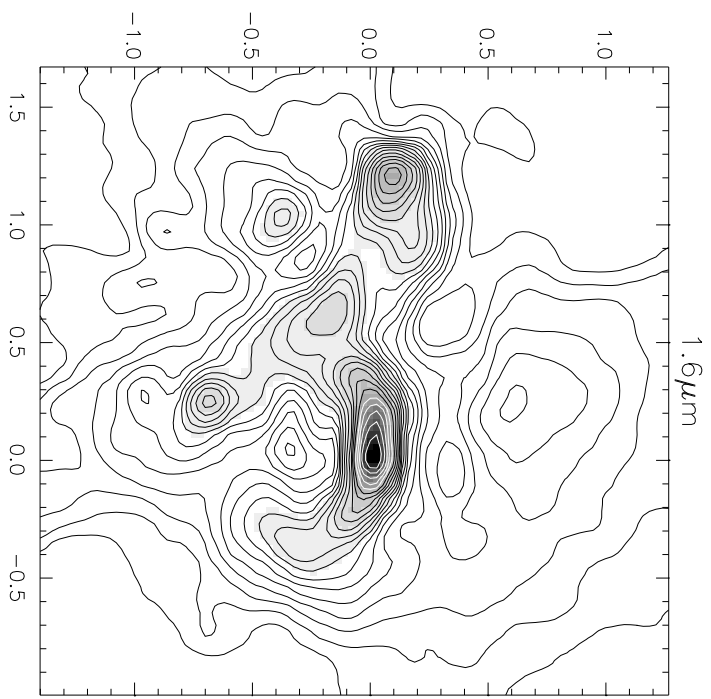
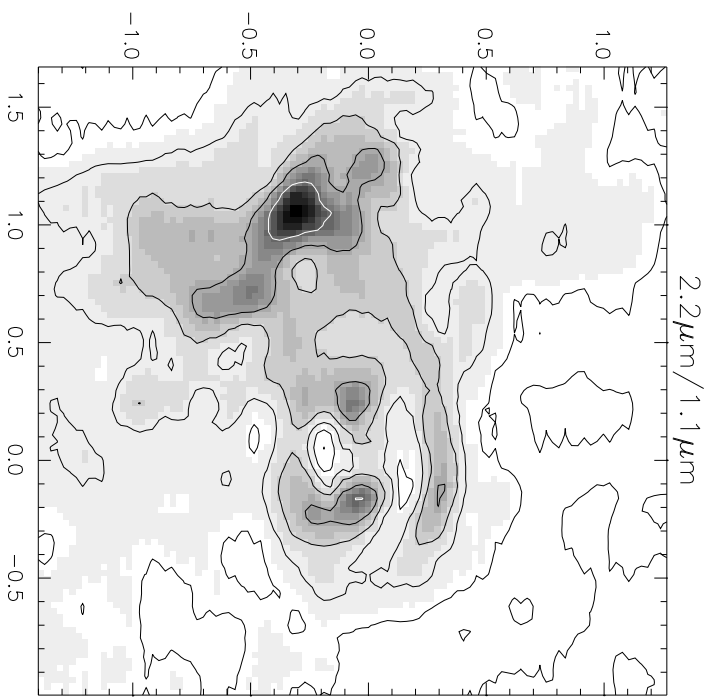
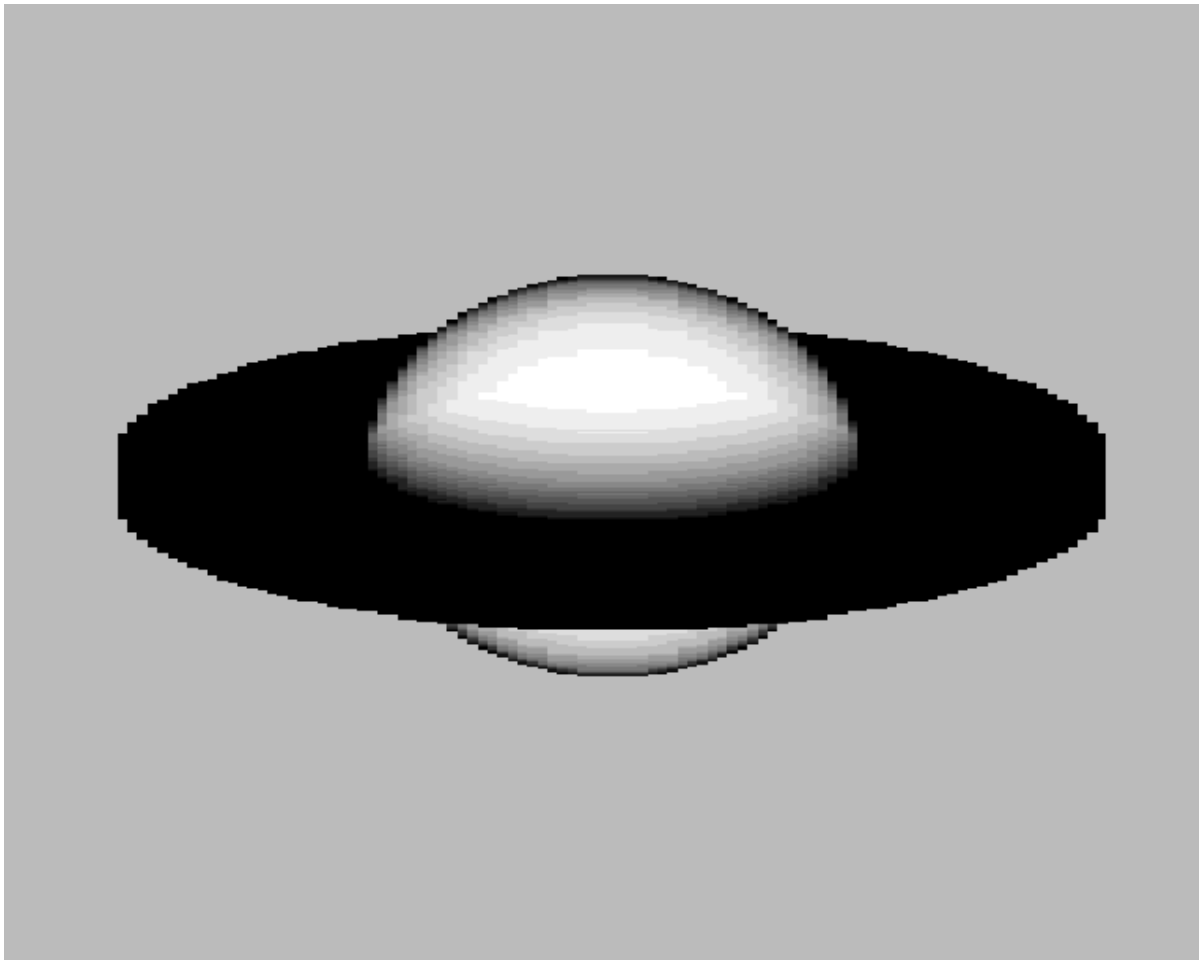
\end{document}